\documentstyle[12pt]{article}
\newcommand{\pbp}{\langle \bar \chi \chi \rangle}

\def\tr{\mathop{\rm tr}\nolimits}
\title{Lattice QCD with 8 Light Quark Flavors\thanks{This
research was supported in part by the U.S. Department of
Energy.}}
\author{Frank R.~Brown\thanks{Present address, The First Boston
Corporation, Park Avenue Plaza, NY, NY~10055.},
Hong Chen\thanks{Present address, IBM Research, P.O.~Box 218,
Yorktown Heights, NY~10598.}, Norman H.~Christ, \\
Zhihua Dong, Robert D.~Mawhinney, Wendy Schaffer, \\
and Alessandro Vaccarino$^\fnsymbol{footnote}$
     \\
     \\
     Columbia University \\
     Department of Physics \\
     New York, NY~10027, USA}
\begin{document}
\maketitle

\begin{abstract}
     QCD with eight flavors is studied on $16^3\times N_t$ lattices with
$N_t=4$, 6, 8, 16 and 32, a dynamical quark mass $ma=0.015$ and
lattice coupling $\beta=6/g^2$ between 4.5 and 5.0.
For $N_t=16$ and 32, hadron masses and
screening lengths are computed for a variety of valence quark
masses.  The previously observed, strong, first-order transition for
$N_t=4$, 6 and 8 is seen, for $N_t=16$, to become a
$\beta$-independent, zero-temperature transition characterized by
a factor of $\approx 3$ change in lattice scale.  This
strong, first-order transition restores chiral symmetry, at least
for $N_t=4$, 6 and 8, producing a chirally symmetric, weak-coupling
phase.  However, as $N_t$ increases to 16, the chiral symmetry
properties of the weak-coupling side of the zero-temperature transition
are unclear and offer a hint of a normal, finite-temperature,
chiral symmetry breaking transition in the weak-coupling
phase.
\medskip

\end{abstract}

\def\thepage{CU--TP--541}
\thispagestyle{myheadings}
\newpage
\pagenumbering{arabic}
\addtocounter{page}{1}

\section{Introduction}
\label{sec:introduction}

     Among the possibilities offered by the numerical simulation of
Quantum Chromodynamics (QCD) is that of studying a variety of
values of the physical parameters of the system.  In particular, a
deeper understanding of the physics of QCD may result from studying
its dependence on the dimension of the fundamental representation
of the gauge group (the number of colors, $N_C$) and the number of
quark flavors, $N_f$.  In this paper we pursue the later
alternative by considering the case of QCD with eight light quark
flavors.

     By considering a range of lattice shapes and couplings we
intended to study the effects of eight quark flavors on both the
finite-temperature QCD phase transition as well as the hadron
spectrum at zero temperature.  However, as we will see, such an
investigation of the hadron spectrum is seriously impeded by the
complex phase structure of eight-flavor QCD.

     Our work extends earlier $T > 0$, eight-flavor
calculations\cite{KOGUT_PRL,KOGUT_NP,FUKUGITA,OHTA}
to larger lattices, smaller quark masses and smaller
lattice spacings, with perhaps surprising results.  The earlier work
on $8^3 \times 4$\cite{KOGUT_PRL,FUKUGITA}, $6^4$\cite{KOGUT_NP},
$8^4$\cite{KOGUT_NP,FUKUGITA} and $16^3 \times 4$ and $\times
6$\cite{OHTA} lattices shows a quite strong, first-order transition.
Comparing the transition for $N_f=2$, 3, 4 and these eight-flavor
calculations reveals a strengthening of the transition as the
number of flavors increases\cite{OHTA,GOTTLIEB}. Both greater
metastability at the critical value of $\beta$ and an expanded
window of small quark mass within which the transition occurs are
found as $N_f$ is increased.  Regardless of the number of
flavors, significant variation of $\beta_c$ is
seen as the number of sites in the temperature direction, $N_t$, is
increased from 4 to 6, as is expected for a ``finite-temperature''
transition.

     As we show below, this pattern changes significantly for
larger lattices with $N_f = 8$.  First, the variation of $\beta_c$ with
$N_t$ has vanished for $N_t \ge 8$;  both $16^3 \times 8$ and $16^3
\times 16$ lattices show a strong, first-order transition at the
same value of $\beta$.  Thus the temperature dependent transition
seen for $N_t < 8 $ has become a temperature independent
``bulk'' transition for $N_t \ge 8$, which we describe as separating
``strong''- and ``weak''- coupling phases of the eight-flavor theory.
(The presence of a bulk transition for large $N_t$ is suggested by
earlier $N_t$-independent jumps seen in $N_f=8 $
calculations with heavier quarks.\cite{KOGUT_NP})
Second, although the valence quark mass dependence of $\pbp$ suggests
that the transition seen for
$N_t=4$, 6 and 8 is one of chiral-symmetry restoration (as
is the case for $N_f = 2$, 3 and 4), the situation is
less clear for $N_t=16$, where $\pbp$ shows non-linear behavior for
small valence quark mass in the weak-coupling phase.

     However, if $\beta$ is increased to $\approx 0.3$ above
$\beta_c$ for the bulk transition, chirally symmetric behavior is seen
for both $\pbp$, which now depends linearly on the valence quark mass,
and the hadron correlation functions.  This suggests that in addition
to the first-order, bulk transition separating strong- and weak-
coupling phases, the weak-coupling phase may itself be divided into
two distinct phases separated by a normal, finite-temperature
(i.e., $N_t$ dependent) transition or cross-over region.  In fact,
the non-linear behavior seen for $\pbp$ in the weak-coupling phase
is reminiscent of the mass dependence of $\pbp$ seen in four-flavor
calculations on $10^3 \times 6$ when $\beta \approx
\beta_c$\cite{DETAR_KOGUT}.

     This behavior is described by the phase structure in the
$\beta-N_t$ plane shown in Figure~\ref{fig:phase}.  (The figure
depicts a lattice of infinite spatial extent.) The solid line
represents the strong, first-order transition, which varies
with $N_t$ for $N_t<8$ and becomes a zero-temperature or bulk
transition for $N_t \ge 8$.  The dashed line expresses our
speculation that a normal, finite-temperature transition or cross-over
region is also present in the weak-coupling phase.  This line
is drawn with a slope given by the perturbative renormalization
group, since this gives the dependence of $\beta_c$ on $N_t$ as
$N_t \rightarrow \infty$ for a physical transition/cross-over.
The dotted line is an extension of the dashed line to values of
$\beta$ smaller
than $\beta_c$ for the bulk transition, since the strength of
the bulk transition and the properties of our updating algorithm
make it possible
to study the weak-coupling phase when it is only metastable.
The solid squares locate parameter values that we have
studied and the open squares mark values of $\beta_c$ that have
been identified.

     Unfortunately, a study of $T=0$ hadron masses in the
weak-coupling region must be performed within the wedge-shaped region
in Figure~\ref{fig:phase} that is bounded below by the dashed line
and to the left by the solid line.  As will be discussed,
volumes larger than $16^3$ will be required to unambiguously
recognize this region, let alone perform a meaningful mass calculation
there.

In Section~\ref{sec:calculation} we describe the calculations that
have been performed and the methods used in both generating
the gauge configurations studied and constructing the various
observables computed.  The transition for $N_t=4$, 6 and 8 is
considered in Section~\ref{sec:n_t=468}, where we present evidence that
it separates chirally symmetric and asymmetric phases.  In
Section~\ref{sec:n_t=16&32} the bulk transition, isolated on $16^3
\times 16$ and $16^3 \times 32$ lattices, is discussed while in
Section~\ref{sec:high_temp} we consider the weak-coupling phase on
$16^3 \times 32$ volumes for larger $\beta$ and hence higher
temperature.  In Section~\ref{sec:n_f-dependence} we discuss the
relation between the phase structure presented here for eight
flavors with that seen for smaller numbers of flavors.  We suggest
that the familiar cross-over region separating strong and weak
coupling in the zero-flavor theory strengthens
with increasing $N_f$ and becomes our $N_f=8$ bulk
transition.  Finally in Section~\ref{sec:conclusion} various
concluding remarks and speculations are presented.

\section{Description of the Calculation}
\label{sec:calculation}

     We have carried out a Monte Carlo evaluation of the
Euclidean-space, Feynman path integral for full QCD using the {\it
R} algorithm of Gottlieb, {\it et al.}\cite{R_ALGORITHM}.  The
calculation required about five months on the 256-node Columbia
machine, a $16 \times 16$ mesh of fast array-processors which
achieves a sustained performance for these calculations of 6.4
Gflops\cite{256_NODE}. The {\it R} algorithm evolves the gauge
fields according to the action
\begin{equation}
{\cal S} = -{1 \over 3}\beta \sum_{\cal P}Re \tr U_{\cal P}
       -{1 \over 4} N_f \ln \det[(D+ma)(D^\dagger+ma)].
\label{eq:action}
\end{equation}
Here the first term is the usual Wilson action with $U_{\cal P}$
the product of the four $SU(3)$ link matrices that border the
plaquette ${\cal P}$.  The second term represents the effects of
$N_f$ degenerate flavors of dynamical fermions of mass $m$.  The
factor of 1/4 preceding this term compensates for the
fermion doubling present in the staggered Dirac operator $D$.
The additional doubling introduced in Eq.~(\ref{eq:action}) by
squaring the Dirac operator is removed by restricting the squared
operator to even lattice sites.  The operator $D$ can be defined by
its action on an $SU(3)$-triplet field $\phi$:
\begin{equation}
(D\phi)_n = {1 \over 2} \sum_\mu
            \eta_{n,\mu}(U^\dagger_{n,\mu}\phi_{n+\mu}
                         -U_{n-\mu,\mu} \phi_{n-\mu}).
\end{equation}
Here $U_{n,\mu}$ and $\eta_{n,\mu}$ are the link matrix and
staggered fermion sign factor associated with a lattice link
extending from the site $n$ in the direction $\mu$ to the site
$n+\mu$.

     In simulating eight flavors of staggered fermions, we have a
choice between two well established methods: the {\it R} algorithm,
which contains finite time-step errors of order $(\Delta \tau)^2$,
and the exact Hybrid Monte Carlo method of Duane, {\it et
al.}\cite{DUANE}, which requires twice the number of Dirac
propagator inversions per unit of Monte Carlo time when used to
simulate $N_f=8$.  We chose the {\it R} algorithm with a time step
$\Delta \tau = 0.0078125$ for this exploratory calculation both
because code to perform the exact, eight-flavor update was not ready
when we wished to begin the simulation and we wanted to reduce the
required computer time.  We have explicitly studied the effects of the
finite time-step errors and made a comparison with results from the
Hybrid Monte Carlo algorithm in our determination of $\beta_c$ for
the transition on $16^3 \times 8$ and $16^4$ lattices as is
discussed in Sections~\ref{sec:n_t=468} and \ref{sec:n_t=16&32}.
The comparisons presented there show the expected quadratic
dependence on $\Delta \tau$.  Although the $\Delta \tau$-errors
found are quantitatively large ({\it e.g.}  5\% in $\beta_c$), the
qualitative features of the calculation appear unaffected.

     Thus, except where otherwise noted, this calculation is
performed with a time step $\Delta \tau = 0.0078125$ and a molecular
dynamics trajectory of length 0.5 time units.  We have used three
types of starting configurations in this calculation: hot starts
where the gauge fields are disordered, cold starts where all the
gauge link matrices are unit matrices and mixed starts which are
described in detail in Section~\ref{sec:n_t=468}.   After each
trajectory, the molecular dynamics ``momenta'' are randomized and
measurements on the link variables carried out.  In particular,
after each trajectory we compute average values of the Wilson
action and the fermion operator $\pbp$.  Our gauge action
is $\langle 1 - 1/3 Re \tr U_{\cal P} \rangle$ and our convention
for $\pbp$ is
\begin{equation}
\pbp \equiv {1 \over 3} {1 \over N_s^3 N_t} \sum_{n} \langle \bar
 \chi_{n} \chi_{n} \rangle,
\label{eq:psibarpsi}
\end{equation}
where the sum is over all points in the lattice.  $\pbp$ is
estimated by
\begin{equation}
\begin{array}{ccl}
 \pbp & = & {1 \over 3} {1 \over N_s^3 N_t} \langle \langle \sum_{l,n}
	    h_l \left( \frac{1}{D+m}
	    \right)_{l,n} h_n \rangle \rangle \nonumber \\
      & = & {1 \over 3} {1 \over N_s^3 N_t}  \langle \langle \sum_{l,n}
	    h_l \left( \frac{m}{D D^\dagger
	    + m^2} \right)_{l,n} h_n \rangle \rangle.
\end{array}
\label{eq:estimator}
\end{equation}
where for each site $n$, $h_n$ is an independent, complex three-vector
of Gaussian random numbers and $\langle \langle \cdots \rangle \rangle$
denotes an average over gauge fields and the random three-vectors,
$h_n$.  For the work in this paper, we have used three sets of $h_n$'s
for each gauge configuration. Also, we restrict the $h_n$'s to
even sites in evaluating the squared operator in
Eq.~(\ref{eq:estimator}) and multiply the result by two.

     Hadron propagators are calculated every 5 units of microcanonical
time from quark propagators determined using Coulomb gauge wall sources
whose spatial size is the spatial volume of the lattice.  The quark
propagator from a source at time slice $t$ is calculated, for each
color index $a=1$--3, using as a source an $SU(3)$ triplet field
$h^b_{\vec{n},t'}$, given by
\begin{equation}
  h^b_{\vec{n},t'} = \delta_{a,b} \delta_{\bar{n}_{1},0}
    \delta_{\bar{n}_{2},0} \delta_{\bar{n}_{3},0} \delta_{t',t},
\end{equation}
where $\bar{n} = n$ mod 2.
For a given source time slice, the three quark propagators are then
combined into hadron propagators corresponding to matrix elements of
the five conventional local hadron operators\cite{OPERATORS}.  In order
to improve the statistical accuracy of our results, we use the
average-over-time-slice (AOTS) method\cite{HONG} in which this hadron
propagator calculation is performed $N_t$ times placing the wall
source on each time slice in the lattice.  The resulting $N_t$
propagators are then averaged together.

     In both the updating steps and the calculation of the hadron
propagators we must solve a Dirac equation of the form $(D+ma)y=h$.
We perform the required inversion of $D+ma$ using the conjugate
gradient algorithm.  We iterate this method until our approximate
solution after the $i$th iteration, $y_i$, yields an appropriately
small residual vector $r_i=(D^\dagger D+(ma)^2)y_i-(D+ma)h$.
Specifically, we perform the inversion on the even sublattice
(since the solution on the odd sublattice can be found from this)
and iterate until
\begin{equation}
\sqrt{(r_i,r_i)/(h,h)} \le \Delta,
\end{equation}
where the inner product $(a,b)$ of the complex vectors is over
even lattice sites and colors.
For the inversions that occur in the updating steps and $\pbp$ we
use $\Delta=6.38 \times 10^{-5}/ \sqrt{N_t}$ and perform typically
between 300 and 700 conjugate gradient iterations, depending on $\beta$,
for $N_t = 32$.  For the hadron propagator calculation we use the
somewhat more stringent condition $\Delta=2.21 \times 10^{-6}$,
yielding 700 to 800 iterations for $N_t=32$ with $ma = 0.015$.

     Much is to be learned in calculations of this sort by varying
the quark mass used in the simulation.  Perhaps of greatest
interest is the variation of $\pbp$ and $m_\pi$ with quark mass.
A non-zero value of $\pbp$ and a zero value $m_\pi$ in the $m
\rightarrow 0$ limit are both definitive indicators of the
spontaneous breaking of chiral symmetry.  Unfortunately, in the
exploratory calculation reported here only the single value $ma=0.015$
has been used.  However, we have computed the dependence of $\pbp$
and the hadron masses on the quark mass that appears in the quark
propagators that explicitly enter the evaluation of the right-hand
side of Eq.~(\ref{eq:estimator}) and the hadron masses.  We have
only considered the case where all quark propagators used to form
a hadron propagator have the same quark mass.

     Thus we distinguish two quark masses that enter our
calculation: the ``sea'' quark mass $m_{sea}$ that enters the quark
determinant in the path integral and the ``valence'' mass $m_{val}$
which appears in the quark propagators that make up the various
observables.  In this way we can define, for example,
$m_\pi(m_{sea},m_{val})$.  A consistent calculation with $N_f$
flavors of degenerate quarks requires $m_{val}=m_{sea}$.  We might
call the quantity $m_\pi(m_{sea},m_{val})$ a ``quenched''
approximation to the proper quantity $m_\pi(m_{val},m_{val})$.
However, a non-vanishing limit of $\pbp$ as $m_{val} \rightarrow 0$
is never-the-less an indicator of spontaneous symmetry breakdown,
although the observable in question is no longer local.  Likewise
the Goldstone theorem implies that if this quenched $\pbp(m_{sea},
m_{val})$ is non-vanishing in the limit $m_{val} \rightarrow 0$,
then the corresponding $m_\pi(m_{sea}, m_{val})$ will also vanish
in that limit.  Clearly, the limit $m_{val} \rightarrow 0$ provides
us with interesting information about the character of the small
eigenvalues of the Dirac operator.

     In fact, if the masses $m_{sea}$ and
$m_{val}$ are sufficiently small that $m_\pi^2$ and $\pbp$ depend on
them linearly, and if $\pbp$ is non-vanishing for both the limits
$m_{val}=m_{sea} \rightarrow 0$ and $m_{sea}$ fixed, $m_{val}
\rightarrow 0$, then necessarily the two values of $m_\pi^2$ agree:
\begin{equation}
m_\pi^2(m_{sea},m_{val})=m_\pi^2(m_{val},m_{val}) + O(m^2),
\end{equation}
since both sides are linear functions of $m_{val}$ which agree at
two points, $m_{val}=m_{sea}$ and $m_{val}=0$.  Because the
majority of the linear mass dependence of $\pbp$ comes from the
quadratically divergent term proportional to $m_{val}$ we might
also expect this quenched calculation of $\pbp$ to be quite
accurate.  Our $N_f=2$ calculations\cite{HADRON_SHORT}, in which
four values of $m_{sea}$ were used, bear these expectations out.  In
Figure~\ref{fig:2f_extrap} we show $m_\pi^2$ and $\pbp$ for the four
normal points\cite{HADRON_SHORT} $m_{val}a=m_{sea}a=0.01,$ 0.015,
0.02, 0.025 and a fifth, quenched point\cite{HADRON_LONG}
$m_{sea}a=0.01$, $m_{val}a=0.004$.  The values for the quenched
point are $m_\pi a=0.173(5)$ and $\pbp = 0.0157(1)$.  A linear fit to
these five points yields:
\begin{equation}
\begin{array}{rlclr}
m_\pi^2 a^2 & = 0.0033(15)            & + & 5.86(7) \,m_{u,d}a
                                      & (\chi^2/dof=26/3), \\
\pbp a^3    & = 0.00768(11)           & + & 2.034(6) \,m_{u,d}a
                                      & (\chi^2/dof=14/3).
\end{array}
\label{eq:2f_fit}
\end{equation}
Although the $\chi^2$ for these fits is large, one can see
{}from Figure~\ref{fig:2f_extrap} that the $m_{sea}a=0.015$ point is
the dominant contributor to the large $\chi^2$, not the point
at $ma=0.004$.  Leaving out the $m_{sea}a=0.015$ point gives
fits with slopes and intercepts that are the same within errors
and which have $\chi^2/dof = 0.2/2$ and 0.5/2, respectively.

\section{Chiral-symmetry Restoration: $N_t$ = 4, 6 and 8}
\label{sec:n_t=468}

     As noted in the introduction, previous studies on lattices
with $N_t \leq 8$ have observed a strengthening of the chiral
transition for an increasing number of quark flavors.  For $N_f =
8$, $\pbp$ has been seen to change by about a factor of two across
the transition for quark masses of 0.1 on lattices with $N_t = 4$,
6 and 8\cite{KOGUT_NP,OHTA}. Previous eight-flavor work did not include
an extrapolation of $\pbp$ to zero quark mass in the weak-coupling
phase to demonstrate that the transition restored chiral symmetry.
In this section we report on our investigation of this transition
for $N_t=4$, 6 and 8 lattices---both the accurate determination of
$\beta_c$ and the $m_{val} \rightarrow 0$ extrapolation of $\pbp$
which gives evidence that this transition does restore chiral
symmetry.  Our $N_t=4$ results were obtained from a series of
simulations described in Table~\ref{tab:run_4} while Tables
\ref{tab:run_6} and \ref{tab:run_8} contain a similar description
of the $N_t=6$ and 8 runs. All the lattices had a spatial volume
of $16^3$.

     The accurate determination of $\beta_c$ for a strong,
first-order transition presents a familiar dilemma: if we work with a
large spatial volume, the considerable metastability of both phases
implies a large range of $\beta$ within which each phase appears to
be stable, even for quite long Monte Carlo evolution times. (At
least if we use existing, local updating algorithms.)  If a
sufficiently small spatial volume is used to eliminate this
metastability, the transition may be significantly distorted by
finite-volume effects.  We solve this problem by working with a
large volume but beginning with a configuration in a mixture of
phases.  Starting with such a mixed phase, very small changes in
the choice of $\beta$ cause the system to rapidly evolve into
either of the two phases\cite{MIXED}.

     Our results for $\beta_c$ and $\pbp$ at $N_t=4$, 6 and 8 have
been obtained by starting from mixed-phase configurations
generated as follows:  An initial $\beta$ value was chosen for
which hot and cold starts gave two metastable phases, with different
values of $\pbp$ and gauge action, which were stable for 50 or more
units of microcanonical time.  We then chose a configuration from
one phase and evolved it, changing $\beta$ every 10--20 units of
time until the configuration had a value of $\pbp$ and action close
to halfway between their values in the two phases. (Both $\pbp$ and
the action reached their halfway values concurrently.)  The final
value of $\beta$ used in evolving this mixed-phase configuration
was chosen to keep both $\pbp$ and the action roughly constant for
10--20 time units, thus diminishing any ``inertia'' the lattice
might have pushing it in the direction of either phase.

     The upper curve in Figure~\ref{fig:mix1_nt4} shows the
evolution of $\pbp$ produced when generating a mixed-phase
configuration from a hot start for an $N_t=4$ lattice.  The first
80 time units show normal evolution with
$\beta=4.5$.  During the next 95 time units $\beta$ was continually
adjusted to produce the intermediate value of $\pbp$ shown.

     After generating a mixed-phase configuration, $\beta_c$ was
found by performing a series of evolutions, with different values
of $\beta$, each starting from the given mixed-phase
configuration.  An example of this procedure is represented in
Figure~\ref{fig:mix1_nt4_beta}.  There we show the evolution of
$\pbp$ for five values of $\beta$, starting from the mixed-phase
configuration generated from an initial hot start.  Clearly
$\beta=4.5$ lies on the strong-coupling side of the transition
while $\beta=4.65$ falls on the weak-coupling side.  The slow
evolution of the $\beta=4.58$ run locates $\beta_c$ while the short
runs at 4.55 and 4.6 suggest $\beta_c=4.58(1)$ as a reasonable
conclusion for $\beta_c$ with errors.

     In order to demonstrate the reliability of this procedure we
repeated the determination using a second, mixed configuration
generated instead from an initial cold start.  The evolutions for
three choices of $\beta$ shown in Figure~\ref{fig:mix2_nt4_beta}
behave in a manner very consistent with
Figure~\ref{fig:mix1_nt4_beta} and the value $\beta_c=4.58(1)$
deduced above.  This indicates that our mixed-phase configurations
are independent of whether they were made from a hot or cold start
and argues against any bias toward one phase.  In addition, we have
never seen an evolution beginning from one of our mixed-phase
configurations begin to change in the direction of one phase and
then reverse itself.  This indicates that our mixed-phase
configurations have no inertia toward a particular phase.

     The series of evolutions used to determine $\beta_c$ for
$N_t=8$ is shown in Figure~\ref{fig:mix_nt8_beta}, from which we
deduce $\beta_c=4.73(1)$.  Although not shown, the evolutions of
the action are very similar to those for $\pbp$.
Table~\ref{tab:betac} lists our results for $\beta_c$ for $N_t =
4$, 6 and 8.  For comparison this table also includes the $N_t=16$
results which are discussed in the next section.

     Given the ease with which we can determine $\beta_c$ and its
importance in our later considerations, it is reasonable to study
the effects of finite time-step errors by computing $\beta_c$ for
a number of choices for $\Delta \tau$.  We concentrated on the
$N_t=8$ case with the results given in Table~\ref{tab:betac}.  As
is indicated in Table~\ref{tab:run_8} these values of $\beta_c$
were determined by the procedure described above, starting from the
same mixed phase (generated with $\Delta \tau = 0.0078125$) and
evolving using a series of updating schemes with $\Delta \tau=
0.002$, 0.005, 0.0078125 and 0.0125 and with the exact Hybrid Monte
Carlo algorithm.

     Figure~\ref{fig:betac} shows the dependence of $\beta_c$ on
$\Delta \tau$.  Leaving out the point at $\Delta \tau = 0.0125$ and
the point from the exact algorithm, we find that $\beta_c$ is fit by
\begin{equation}
   \beta_c(\Delta \tau) =  4.58(1) + 2460(250) \, (\Delta \tau)^{2}
\end{equation}
with $\chi^2/dof=0.02/1$.  (The fit for $\beta_c$ using the points up
to 0.0078125 is good enough to make it hard to fit the 0.0125
point, even if higher order terms are included.)  Clearly, we find
the expected $(\Delta \tau)^{2}$ dependence of $\beta_c$ and good
agreement between the constant term in the fit and the value of
$\beta_c$ from the exact algorithm.

     Finally, let us examine the values of $\pbp$ obtained from these
various runs and their dependence on $m_{val}$.  Tables
\ref{tab:result_4}, \ref{tab:result_6} and \ref{tab:result_8} give
$\pbp$ for the masses used. The thermalization times, $\tau_{eq}$,
given in these three tables are estimates obtained by eye from the
plots of $\pbp$ and vary because the thermalization time depends on
$\beta - \beta_c$, which is not constant for the different runs.

     As can be seen in Figure~\ref{fig:468extrap}, $\pbp$
extrapolates linearly to zero as $m_{val} \rightarrow 0$ for
$N_t=4$, 6 and 8 on the $\beta > \beta_c$ side of the transition.
The fits are forced through the origin and have $\chi^2/dof = 0.39/1$,
3.9/2 and 5.1/1, respectively.  The figure clearly shows that our
results
are consistent with a chirally symmetric, weak-coupling phase.
In addition, for $N_t=8$ we can check that similar, chirally symmetric
behavior is seen as the time step is varied by examining $\Delta
\tau = 0.0125$ and 0.005. For these cases we also find good linear fits
for $\pbp$ as a function of quark mass.  When forced through zero,
the fits have $\chi^2/dof = 0.74/1$ and 0.007/1, respectively.

\section{Evidence for a $T$=0 transition}
\label{sec:n_t=16&32}

     The eight-flavor results described in the preceding section and
earlier work of others look much like the chiral-symmetry-restoring
phase transition seen for four flavors of light quarks on lattices
of similar size.  These results are usually interpreted as a
lattice approximation to a phase transition at non-zero temperature
for the continuum field theory.  The variation of the value of
$\beta$ where the transition occurs ($\beta_c$) with $N_t$ supports
this interpretation.  One expects the physical temperature of the
lattice at the critical point,  $T=(N_ta)^{-1}$, to be fixed so
that changes in the lattice spacing $a$ resulting from changes in
$\beta$ must be compensated by changes in $N_t$.  Such behavior is
quite well established in pure QCD where the variation of $a$ (and
hence $N_t$) with $\beta$ predicted by the perturbative
renormalization group is seen\cite{PURE_GAUGE_TRANSITION} on
lattices as large as $24^3\times 16$.

     However, for the eight-flavor transition we find the critical
value of $\beta$ does not change when $N_t$ is increased from 8 to
16.  For $N_t=16$ we continue to see a very strong, first-order
transition even though the $16^4$ lattice now has a spatial size no
larger than the temporal extent.  This apparent space-time volume
independence of $\beta_c$ suggests the transition will persist with
this fixed value of $\beta_c$ even for a system of infinite spatial
and temporal extent.  We conclude that for $N_f=8$ there is a
strong, first-order $T=0$ or bulk transition separating the strong-
$(\beta \le \beta_c)$ and weak- $(\beta \ge \beta_c)$  coupling
regimes.

     Although the strong-coupling phase seen on $16^4$ lattices
appears much like that found for $N_t=4$, 6 and 8, the weak-coupling
phase is different in three respects.  First, the precisely linear
dependence of $\pbp$ seen in Figure~\ref{fig:468extrap} for
$N_t=4$, 6 and 8 as $m_{val} \rightarrow 0$ becomes significantly
non-linear for the $16^4$ lattice.  Second, for $\beta=4.65$ $\pbp$
doubles as $N_t$ is increased from 4 to 16.

     Finally for $m_{val}a=m_{sea}a=0.015$ the hadron spectrum does
not show the degree of parity doubling that might be expected for a
phase in which chiral symmetry has been restored.  For example, we
see nearly exact parity doubling for larger $\beta$ as is described
in Section~\ref{sec:high_temp}.  In fact, the values of the hadron
masses and $\pbp$ seen here at $\beta=4.65$ are more similar to
those found for two flavors in the low temperature, chirally
asymmetric phase using the same quark mass and lattice size.
However, we do find parity doubling in the limit
$m_{val} \rightarrow 0$.  We tentatively interpret these
results as suggesting that these $\beta=4.65$, $16^3 \times 16$ and
$\times 32$ lattices lie in a transition region that occurs in the
{\em weak}-coupling phase---a finite-temperature transition/cross-over
leading to spontaneous chiral-symmetry violation on lattices of
larger spatial and temporal extent.

     Let us now describe these results in greater detail.  Tables
\ref{tab:run_16} and \ref{tab:run_32} give the particulars of the
$16^3 \times 16$ and $16^3 \times 32$ calculations on which our
conclusions are based.  Again, except where explicitly noted, we
used the inexact R algorithm with the step size $\Delta \tau
=0.0078125$.  Because of the limited length of some of these runs,
the amount of data discarded as not in equilibrium will be
discussed  on a case-by-case basis below.

     Figure~\ref{fig:4_65_two_phase} shows the evolution of $\pbp$
starting from hot and cold starts for $\beta=4.65$ on a $16^3
\times 16$ lattice.  The persistence of two phases over more than
300 time units, a time scale considerably greater than the initial
thermalization time of $\le 50$ time units, is evidence for a
strong, first-order transition.  As further evidence for a
first-order transition, Figure~\ref{fig:4_60_tunnel} shows an evolution
for the $16^3 \times 32$, $\beta=4.60$ cold-start run in which
a tunneling event occurs at $\tau \approx 250$ time units,
suggesting that the weak-coupling phase becomes unstable as $\beta$
is decreased from 4.65 to 4.60.

     The critical coupling $\beta_c$ on the $16^4$ lattice is
determined by the same procedure described earlier for $N_t=4$, 6 and
8.  We created a ``mixed'' start by beginning with the weak-coupling
configuration whose evolution is shown in the lower curve in
Figure~\ref{fig:4_65_two_phase} and then varying $\beta$ by hand
for 55 time units to obtain a configuration with a value of
$\pbp$ lying midway between the strong- and weak-coupling values
seen in the figure, {\it i.e.} $\pbp \approx 0.2$.  This final
configuration is then used as the beginning for the three different
runs shown in Figure~\ref{fig:mix_nt16_beta}.  This figure
establishes $\beta_c=4.73(1)$, precisely the result found above for
$N_t=8$.  Because of the significant time-step dependence seen
earlier for $\beta_c$, we carried out this procedure a second time
for $\Delta \tau =0.005$ and determined for that case,
$\beta_c=4.62(1)$, again in agreement with the $N_t=8$ result for
that smaller time step.  We conclude that this strong, first-order
transition has become independent of the lattice size for $N_t \ge
8$ and hence is a $T=0$ transition.

     The lack of dependence of $\beta_c$ on $N_t$ seen for $N_t \ge
8$ is quite consistent with the $N_t$ dependence of the
discontinuity in the gauge action across the transition.  For a
normal, finite-temperature transition, an increase of $N_t$  by a
factor of 2 would correspond to a decrease of the lattice spacing
$a$ by a factor of 2.  Since for such a transition the
discontinuity in the action is proportional to a physical latent
heat, the jump in the action should decrease by a factor of
$2^4=16$ when $N_t$ increases from 8 to 16.  In fact a similar
factor of $(4/6)^4$ is seen for the $N_f=4$ latent heat when $N_t$
is increased from 4 to 6\cite{THERMO_NF4}.  However, our $16^3
\times 8$ and $16^4$ results given in Tables~\ref{tab:result_8} and
\ref{tab:result_16} show a large 20\% jump in the action which
changes relatively little between $N_t=8$ and 16.  In particular,
for $N_t=8$ the decrease in the action between $\beta=4.70$ and 4.75 is
0.1202(5) while for $N_t=16$ the difference between the action in the
two metastable phases at $\beta=4.65$ is 0.1150(2)---a decrease by
5\% not by a factor of 16.

     Next let us consider the chiral-symmetry properties of the two
phases separated by this transition.  We have computed $\pbp$ and
the hadron spectrum in each phase for a number of valence quark
masses.  We work at $\beta=4.65$, where there are two
metastable phases as shown in Figure~\ref{fig:4_65_two_phase}.
Although this choice of $\beta$ is below the critical value
$\beta_c=4.73$, at this $\beta$ the weak-coupling phase shows no signs
of instability either during the 342.5 time unit $16^4$ run
(Figure~\ref{fig:4_65_two_phase}) or the 865 time unit run on a
$16^3 \times 32$ lattice.  These results, together with the $\beta=5.0$
results described in the next section are given in
Tables~\ref{tab:result_16}, \ref{tab:result_32},
\ref{tab:masses_sea} and \ref{tab:masses_val_4_65}.

     The second columns of Tables~\ref{tab:masses_sea} and
\ref{tab:masses_val_4_65}, for the strong-coupling phase, were
obtained on a $16^4$ lattice from the microcanonical time range
210---380.  The fitting was done by combining these results into
blocks of 5 time units and the masses came from two parameter fits
assuming a single propagating state.  The $\pi$ mass was determined
by fitting time separations 2 to 8 while the masses for the other
states were obtained from time separations 1 to 5.  The resulting
$\pi$ masses appear to be well determined from the range of time
separations available.  However, the correlators for the other
strong-coupling states decrease so rapidly with separation that
useful information comes only from a few time separations.  As a
result, we are less certain that those masses have taken on truly
asymptotic values.

     For the weak-coupling phase, given in the third
column of Table~\ref{tab:masses_sea} and columns three through six
of Table~\ref{tab:masses_val_4_65} we use the longer $\beta=4.65$
run on a $16^3 \times 32$ lattice discarding the first 382.5 time
units for equilibration.  With this larger time dimension, stable
results are obtained for a larger number of masses.  We follow a
procedure to extract the masses similar to that used
earlier\cite{HADRON_SHORT}: the $\pi$ mass is obtained from a
one-propagating-state, two parameter fit while the other masses come
{}from a two-propagating-state, four parameter fit.  We used a range
of fitting separations, $t_{min} \le t \le 16$ as follows:  For the
quark masses 0.004, 0.01 and 0.015 and the $N$ and $N'$ states we
used $t_{min}=8$ and for all other states $t_{min}=10$.  For the
quark masses 0.025 and 0.05 we used $t_{min}=10$ for the $\pi$
state and $t_{min}=8$ for the others.  With these fitting ranges we
obtain $\chi^2/dof<2$.  All the masses given in
Tables~\ref{tab:masses_sea} and \ref{tab:masses_val_4_65} came from fits
obtained by minimizing $\chi^2$ computed from the full covariance
matrix.  Errors were determined with the jackknife method using
blocks of 15 time units (5 time units for the shorter 0.025 and
0.05 runs) with corrections for autocorrelations in Monte Carlo
time.

     As can be seen from Table~\ref{tab:masses_sea} the strong- and
weak-coupling phases have very different values for the hadron
masses. The masses in the weak-coupling phase are lighter by about
a factor of three, except the pion which we discuss in detail below.
The vacuum expectation value $\pbp$ also
becomes much smaller moving from strong to weak coupling.  However,
by itself, such a jump does not imply that the transition
restores chiral symmetry. For example, naive scaling arguments
would suggest that for small quark mass $\pbp$ should decrease by
a factor of $3^3$ when the hadron masses decrease by a factor of 3.

     To study the chiral symmetry of these two phases we show the
linear extrapolations of $\pbp$ and $m_\pi^2$ as $m_{val}
\rightarrow 0$ in Figures~\ref{fig:4_65_strong_extrap} and
\ref{fig:4_65_weak_extrap}.  The behavior of $\pbp$ and $m_\pi^2$ in
the strong-coupling phase (Figure~\ref{fig:4_65_strong_extrap}) is
easily interpreted.  We are seeing the usual consequences of
spontaneous chiral-symmetry breaking---behavior quite similar to that
shown for $N_f=2$, $\beta=5.7$ in
Figure~\ref{fig:2f_extrap}\cite{HADRON_SHORT,HADRON_LONG}.

     However, the chiral properties of the weak-coupling, $N_f=8$
phase are more ambiguous.  Our earlier $16^3 \times 32$
results\cite{HADRON_SHORT,HADRON_LONG} for $N_f=2$, $\beta=5.7$ and
$m_{sea}a=0.015$ are reproduced in column five of
Table~\ref{tab:masses_sea},
allowing easy comparison of the chiral-symmetry breaking found in
these two sets of spectra.  Although the $m_\pi-m_\sigma$ and
$m_\rho-m_{A_1}$ splittings are significant for the eight-flavor
case, they are perhaps half the size of those seen in the $N_f=2$
case.  In the earlier $N_f=2$ calculation we found $\pbp=0.0385(1)$
so the possible measure of chiral-symmetry breaking,
$\pbp/m_\rho^3$, is the same between the two calculations up to the
20\% level.

     It is also of interest to ask how $\pbp$ changes in the
weak-coupling phase as lattice size increased.  We can directly compare
the $N_t=4$ and $N_t=16$ results for $\beta=4.65$ recognizing an
increase from 0.0343(3) to 0.0711(4).  Such an increase with
increasing $N_t$ might be interpreted as the onset of
chiral-symmetry breaking as the temperature decreases for fixed
$\beta$.

     However, the notion that the weak-coupling phase seen for
$N_t=16$ and 32 shows spontaneous symmetry breaking is not
supported by the dependence of $\pbp$ or the hadron spectrum on
$m_{val}$.  As shown in Figure~\ref{fig:4_65_weak_extrap} $\pbp$
appears to approach 0 as $m_{val} \rightarrow 0$ while the
corresponding limit of $m_\pi$ is small but non-zero.  As can be
seen, the small $m_{val} \rightarrow 0$ limit of $\pbp$ is
surprisingly non-linear for $m_{val}a$ as small as 0.004 but
certainly appears to vanish.  However, the extrapolation of
$m_\pi^2$ is straight forward,
\begin{equation}
\begin{array}{rlclr}
m_\pi^2 a^2  & = 0.0391(13)  & + & 6.90(6) m_{u,d}a
                                          & (\chi^2/dof=1.9/3), \\
\end{array}
\label{eq:4_65_quad_fit}
\end{equation}
giving $m_\pi^2(0) \sim 30$ standard deviations away from zero.  This is
in contrast with the behavior seen in the $N_f=2$ weak-coupling
phase shown in Figure~\ref{fig:2f_extrap} where, as is shown in
Eq.~(\ref{eq:2f_fit}), the extrapolated $\pi$ mass is consistent
with zero.

     In Figure~\ref{fig:4_65_weak_mass} we show the masses of the
three parity partners $N'-N$, $A_1-\rho$ and $\sigma-\pi$ as a
function of $m_{val}$.  Again even the considerable $\sigma-\pi$
splitting disappears linearly as $m_{val} \rightarrow 0$.
Quantitatively, fitting to the three smallest values of $m_{val}$
shown in the plot we obtain:
\begin{equation}
\begin{array}{rlclr}
m_\pi a    & = 0.218(5)  &+&10.7(4)\,m_{val}a  & (\chi^2/dof=2.9/1), \\
\medskip
m_\sigma a & = 0.220(4)  &+&16.5(4)\,m_{val}a  & (\chi^2/dof=5.0/1), \\
m_\rho a   & = 0.359(14) &+&10.8(12)\,m_{val}a & (\chi^2/dof=0.1/1), \\
\medskip
m_{A_1} a  & = 0.371(30) &+&14.1(24)\,m_{val}a & (\chi^2/dof=0.003/1),
  \\
m_N a      & = 0.704(35) &+&11.2(26)\,m_{val}a & (\chi^2/dof=0.01/1), \\
m_{N'} a   & = 0.694(42) &+&13.5(31)\,m_{val}a & (\chi^2/dof=0.08/1), \\
\end{array}
\label{eq:4_65_linear_fit}
\end{equation}
showing detailed chiral-symmetry restoration as $m_{val}
\rightarrow 0$.  Note the linear fit to $m_\pi^2$ in
Eq.~(\ref{eq:4_65_quad_fit}) (using valence quark masses up to 0.050)
and the linear fit to $m_\pi$ in
Eq.~(\ref{eq:4_65_linear_fit}) (using valence quark masses up to 0.015)
give 0.198(3) and 0.218(5) respectively for
$m_\pi(0)$.

     We conclude that the strong-coupling phase seen at
$\beta=4.65$ for $N_t=16$ and 32 shows clear spontaneous violation
of chiral symmetry while the chiral symmetry of the weak-coupling
phase is less obvious.  However, the unusual $m_{val}$ dependence
that we see in the weak-coupling phase for $\beta=4.65$ is very
much like the $m_{sea}$ dependence found earlier in the
four-flavor, $10^3 \times 6$ work of DeTar and
Kogut\cite{DETAR_KOGUT}.  Their results in the critical region
($\beta=5.175$) show a non-linear approach of $\pbp$ to zero as
$m_{sea} \rightarrow 0$.  Likewise, their $\pi$ and $\sigma$
screening lengths, while significantly non-degenerate for
$m_{sea}=0.05$, become equal when extrapolated linearly to
$m_{sea}=0$.  This behavior is precisely the $m_{val} \rightarrow
0$ dependence that we see for these quantities.  Therefore, we
speculate that for $N_t=16$ and $\beta=4.65$, the weak-coupling
phase is itself near a standard, finite-temperature transition
region separating the chirally symmetric, weak-coupling behavior we
see for $N_t \le 8$ from a weak-coupling, chirally asymmetric region
that will be seen for $N_t \ge 16$ on significantly larger spatial
volumes.  This speculation
is represented in Figure~\ref{fig:phase}, where the dashed line
identifies a possible finite-temperature phase transition dividing
the weak-coupling phase into low-temperature, chirally asymmetric
(upper portion) and high-temperature, chirally symmetric (lower
portion) phases.  This line passes near $N_t=16$ at $\beta=4.65$ as
is suggested by our results for these parameter values.

\section{High-temperature Region: $N_t=32$}
\label{sec:high_temp}

     In an attempt to understand the properties of the
weak-coupling phase discussed above for $\beta \approx 4.65$ on $16^3
\times 16$ and $\times 32$ lattices, let us examine a
1325 time unit calculation of $\pbp$ and hadron masses with
$\beta=5.0$ carried out on a $16^3 \times 32$ lattice.
As is discussed below,
we find clear chirally symmetric behavior for this larger value of
$\beta$.  The hadron screening lengths show complete parity
doubling within errors, $\pbp$ extrapolates linearly to zero as
$m_{val} \rightarrow 0$ and $m_\pi$ varies little as $m_{val}
\rightarrow 0$ and has a relatively large $m_{val}=0$ limit.

     These masses or screening lengths were determined from the
evolution interval 605---1325.   The results are shown in
Tables~\ref{tab:masses_sea} and \ref{tab:masses_val_5_00}.  The
fitting procedure is very similar to that used earlier: the
$\pi$-like states were determined from a two-parameter, single-state fit
while the other states from a four-parameter, two-state fit.  For
all masses we used a fitting range from time separations 10 to 16.
The quark masses of 0.01 and 0.015 were analyzed dividing the
data into blocks of 15 time units while the shorter run with the
valence mass of 0.004 used blocks of 5 time units.  In contrast to the
other mass fits discussed in this paper, the $\chi^2$ values were very
large, typically 10 to 30 with 5 degrees of freedom.  However, the
jackknife errors for these $\chi^2$ values were nearly as large
as the $\chi^2$ themselves and
the $\chi^2$ computed ignoring off-diagonal terms in the
correlation matrix are quite reasonable.  We conclude that the fits
are acceptable but that small poorly determined eigenvalues in the
correlation matrix make the determination of $\chi^2$ difficult.

     The hadron spectrum looks very much like that found in earlier
calculations in the plasma phase\cite{DETAR_KOGUT}.  In particular
the masses (or more accurately screening lengths) show remarkable
parity doubling with the parity partners $\pi-\sigma$, $\rho-A_1$
and $N-N^\prime$ having very nearly the same mass.  Likewise we can
examine the extrapolation to zero valence mass of both $m_\pi^2$ and
$\pbp$ shown in Figure~\ref{fig:5_00_extrap}.  Linear fits to the
data yield
\begin{equation}
\begin{array}{rlclr}
m_\pi^2 a^2  & = 0.134(6)  & + & 1.95(52) \,m_{val}a
                                         & (\chi^2/dof=2.0/1), \\
\pbp a^3     & = 0.00052(9) & + & 2.326(7) \,m_{val}a
                                         & (\chi^2/dof=8.2/1).
\end{array}
\label{eq:HIGH_T_EXTRAP}
\end{equation}
In marked contrast
with the behavior seen for $\beta=4.65$, $m_\pi$ depends rather
weakly on the valence quark mass, extrapolating to a value only
10\% below the $m_{val}=0.015$ point, while $\pbp$ extrapolates
to a very small value. Given the non-vanishing of $m_\pi^2$ as
$m_{val} \rightarrow 0$, the statistically non-zero value of
$\pbp$ may reflect the use in Eq.~(\ref{eq:HIGH_T_EXTRAP})
of a fit neglecting the correlations between results for
different valence quark masses.

     Although the behavior seen for $\beta=5.0$ is very clearly
that expected from QCD at finite temperature, we should emphasize
that our $16^3 \times 32$ lattice is awkward to interpret as
representing finite temperature.  The nominal ``temperature''
direction with extent $N_t=32$ and the required anti-periodic
boundary conditions for the fermions is the longest dimension in
the lattice.  Probably the best interpretation of our space-time
volume is as a $16^2 \times 32$ spatial volume with a temperature
dimension corresponding to 16 lattice units.  Clearly the behavior
seen on this $16^3 \times 32$ lattice may show significant finite-volume
distortions relative to a proper, finite-temperature calculation on
a $N_s^3 \times 16$ lattice with $N_s \gg 16$.

     The contrast between the $\beta=5.0$ behavior just described
and the $\beta=4.65$, weak-coupling phase discussed in
Section~\ref{sec:n_t=16&32} supports the hypothesis that for the
$16^4$ lattice, $\beta=4.65$ lies near a transition region.  By
increasing $\beta$ from 4.65 to 5.0 the degree of chiral symmetry
has dramatically increased for fixed $m_{val}=0.015$ and the
non-linear $m_{val}$ dependence of $\pbp$ has disappeared.

\section{Possible $N_f$-Dependence of QCD}
\label{sec:n_f-dependence}

     In this section we present a possible picture of the
$N_f$ dependence of QCD that connects earlier work for $N_f=0$, 2,
3 and 4 with the $N_f=8$ results given here.  Although the picture
described below is supported by the presently available numerical
results, it is far from unambiguously established by our current
calculations.

     We would like to interpret the eight-flavor bulk transition
seen here as an outgrowth of the strong- to weak-coupling cross-over
region seen in pure SU(3) gauge theory for $\beta \approx 5.6$.
The variation seen in this region provides a connection between strong
coupling, where the scale of the physics is controlled by the lattice
spacing, and weak coupling, where the scale is unrelated to the
lattice spacing.  (The width of the cross-over region for SU(2)
seen using the standard Wilson action can be altered by including an
adjoint representation contribution to the
action\cite{SU(2)_MIX_ACTION}.)
As one passes through this region from strong to weak
coupling, the hadronic energy scale (measured in lattice units)
decreases at a rate faster than predicted by the perturbative
renormalization group.  This was seen quite clearly for $T_c\,a$
in pure SU(3) by Kennedy {\it et al.}\cite{SU(3)_SCALE_VIOLATION}.

We hypothesize that adding additional light dynamical quarks to
QCD promotes a rapid cross-over region to a phase transition,
the first-order, bulk transition seen here, and that the addition
of the quarks is similar to the effect of a non-zero adjoint action in
the pure SU(2) case.  The effect of this increasingly sharp cross-over
region on the finite-temperature QCD phase transition might be
deduced from Figure~\ref{fig:squash}.  Here we represent the cross-over
region for a system of infinite space-time volume by the interval
of $\beta$ between the vertical dotted lines.  The more rapid
variation of $N_t$ with $\beta_c$ within this region joins the
relatively large value of $T_c\,a$ for small $\beta$ with a smaller
value for large $\beta$.  If this region narrows as the number of
flavors increases, sharpening into an actual discontinuity, the QCD
phase transition for values of $\beta_c$ within this cross-over region
might be expected to sharpen as well.  Such a sharpening of the
transition as the number of flavors increases is certainly well
established by current simulations\cite{GOTTLIEB}.

     Furthermore, such a narrowing of the cross-over region with
increasing $N_f$ would imply a corresponding increase in slope of
$N_t$ versus $\beta$ within this region.  In
Figure~\ref{fig:n_f-summary} we plot the variation of $N_t$ with
$\beta_c$ seen in simulations for zero\cite{000}, two\cite{222} and
four\cite{444} flavors
together with that seen here for $N_f=8$. For the two and four
flavor cases, linear interpolation has been used to produce
a value for $\beta_c$ at $m_{sea}a=0.015$.  The behavior predicted
by the perturbative renormalization group is shown by the slopes of
the dashed lines in the figure.  This figure is consistent
with the view that temperature dimensions between $N_t=4$ and 8 lie
within this cross-over region and that the slope of the
$N_t$-versus-$\beta_c$ curve is increasing with $N_f$.

     Although far from well established, this picture is nicely
consistent with the eight-flavor results presented in this paper.
If the cross-over region shown in Figure~\ref{fig:squash} shrinks
to a vertical line as $N_f \rightarrow 8$, the finite-temperature
phase transition effectively disappears for the corresponding
interval of $N_t$, being engulfed there by the bulk transition.  An
eight-flavor phase diagram very much like that shown in
Figure~\ref{fig:phase} results, in a manner that might be described
as follows:

\begin{enumerate}
\item For strong coupling and small values of $N_t$ one expects to
see a single phase transition that separates very different strong-
and weak-coupling regimes.  $N_t$ should vary with $\beta$ for
small values of $N_t$ characteristic of the small length scale
important at strong-coupling ($4 \le N_t \le 8$ in
Figure~\ref{fig:phase}).
\item For values of $N_t$ larger than this strong-coupling length
scale, the transition becomes a bulk transition, with a fixed value
of $\beta=\beta_c$.  Now the dramatic change in hadronic length
scale, which occurred rapidly in the cross-over region for $N_f \le
4$, happens discontinuously across this bulk transition (a scale
change by a factor of 3 in our case for $N_t \ge 8$).  The
finite-temperature, $N_t$-dependent transition has disappeared.
\item An apparently independent finite-temperature transition should
occur in the weak-coupling phase at a much larger value of $N_t$.
This larger value of $N_t$ (determined for $\beta$ near the
bulk transition) should be related to the values of $N_t$
identified in 1 above.
These two scales should be related by the same factor that describes the
jump in the length scale of hadronic phenomena across the bulk
transition.  Thus, in our case we might expect a weak-coupling,
finite-temperature phase transition to occur for $\beta = 4.73 $ and
values of $N_t$ in the range of 3 (the jump in hadronic length scale)
$\times 8$ (the $N_t$ where the transition becomes $N_t$ independent).
In fact, as discussed in Section~\ref{sec:n_t=16&32}, we have some
evidence for such a weak-coupling, finite-temperature transition for
$\beta=4.65$ and $N_t \approx 16$.  A choice of $\beta = 4.73$ and
$N_t=20$ is used to locate the dashed curve in Figure~\ref{fig:phase}.

We can compare the ratio $T_c/m_\rho$ for this conjectured
finite-temperature transition with the value for other numbers of
flavors.  For zero, two and four flavors, extrapolated to zero
quark mass, $T_c/m_\rho = 0.26$, 0.19 and 0.13\cite{GOTTLIEB}.
Using $m_\rho$ at $\beta=4.65$ and assuming a monotonic decrease
in $T_c/m_\rho$ with the number of flavors, we find $N_t \ge 21$.

\end{enumerate}

Of course, a $T=0$ or bulk transition for which $\beta_c$
becomes precisely independent of $N_t$ should separate two
phases each of whose properties are independent of $N_t$.  This is not
the case for a quark-gluon plasma, a natural candidate for the high
temperature, weak-coupling phase represented by the lower right
region of Figure 1.  However, the $1/N_t^4$ behavior expected for the
free energy of a quark-gluon plasma becomes sufficiently weak for
$N_t \ge 8$ as to be completely consistent with the $N_t$-dependence
of $\beta_c$ that we see.  In fact, $\beta_c(N_t)$ for our four values
of $N_t$ is well fit by
\begin{equation}
\begin{array}{rlclr}
   \beta_c(N_t) & =  4.737(6) & - & 40(3)/N_t^4 & \chi^2/dof=0.6/2.
\end{array}
\end{equation}

A final implication of our hypothesis, in analogy with the pure SU(2)
case, is that for large $N_t$, the strong- and weak-coupling sides of
the bulk transition may be continuously connected by using an action
which includes single plaquette contributions from higher
representations of SU(3).  This is consistent with our picture that the
bulk transition, for large enough lattices, is between two chirally
asymmetric phases.

\section{Conclusion}
\label{sec:conclusion}

     Our $N_f=8$ studies are well summarized by the $\beta-N_t$
phase diagram given in Figure~\ref{fig:phase}.   Let us conclude with
the following remarks:

\begin{enumerate}
\item We have argued in Section~\ref{sec:n_f-dependence} that the
phase structure shown in Figure~\ref{fig:phase} may be quite
consistent with the flavor dependence of the QCD phase transition
seen previously for $N_f=0$, 2, 3 and 4.  However, in that
discussion we argued that the well-known strengthening of the
finite-temperature transition for $4 \le N_t \le 8$ that is seen
with increasing $N_f$ came from approaching the strong, $N_f=8$
bulk transition.  Thus this important feature, which dominates
present lattice calculations, may be closely tied to a lattice
artifact.  The true, continuum, $N_f$ dependence of the QCD phase
transition may be quite different and may be seen only on much
finer lattices.
\item It is interesting to ask if such a bulk transition for eight
flavors may have already been anticipated.  In fact there have been
a number of papers that have explored possible phase structures for
QCD with a large number of quark flavors\cite{BANKS&ZAKS,KOGUT_NP}.
Surely interesting new behavior is to be expected as one approaches
$N_f=16.5$ where asymptotic freedom is lost.  However, these
theoretical studies typically predict a transition between a
strong-coupling, chirally asymmetric phase with particle-like
bound states and a zero-temperature, weak-coupling phase with Greens
functions showing fractional anomalous dimension and lacking a particle
interpretation.

This behavior is not seen in the weak-coupling phase of
Figure~\ref{fig:phase} for either $\beta=4.65$ or 5.0.
The masses or screening lengths given in Tables~\ref{tab:masses_sea},
\ref{tab:masses_val_4_65} and \ref{tab:masses_val_5_00} are non-zero
and come from fitting the correlation function to a function with
{\em exponential} time dependence.  Thus it is most natural to
interpret these mass results as describing interacting, particle-like
states with definite, non-zero energy eigenvalues.  We believe that the
high-temperature, weak-coupling phase seen in our calculations is
quite conventional, very much like normal, high-temperature QCD seen
for $N_f \le 4$.
\item A further argument for the finite-temperature transition
represented by the dashed line in Figure~\ref{fig:phase} is based
on the 't~Hooft anomaly conditions\cite{T'HOOFT}.  Since the 't~Hooft
anomaly conditions are inconsistent with a chirally symmetric
eight-flavor, color $SU(3)$ theory at zero temperature, we expect that
the chirally symmetric phase to the right of the bulk transition in
Figure~\ref{fig:phase} cannot extend to zero temperature in the
continuum limit.  The finite-temperature transition suggested by
the dashed line in Figure~\ref{fig:phase} insures that this
chirally symmetric phase is restricted to a region of non-zero
temperature in the continuum, $N_t \rightarrow \infty$ limit.
\item Our original objective in undertaking these eight-flavor
calculations was to study the zero-temperature hadron spectrum for
$N_f=8$ to gain some quantitative insight into the effects of the
fermion determinant in lattice QCD, hadron mass calculations.  This
objective has been frustrated by the existence of the $N_f=8$ bulk
transition.

The decrease in $T_c/m_\rho$ expected as $N_f$ increases, requires
that we increase $m_\rho\,a$ by lowering $\beta$ or increase
$1/Ta=N_t$ relative to calculations with smaller $N_f$.  The
$N_f=8$ bulk transition prevents us from increasing $m_\rho\,a$ by
moving closer to the strong-coupling region and forces us to work
instead at larger $N_t$, in particular lattice sizes with spatial
dimensions $\ge N_t \gg 16$.  Thus it appears that both a proper
demonstration of the finite-temperature, weak-coupling phase
transition and such a low temperature study of hadron masses in
eight-flavor QCD forces the use of lattice volumes considerably
larger than $16^3 \times 32$.
\end{enumerate}

After the completion of this work, we became aware of a study of
QCD with many flavors of Wilson fermions in the strong coupling
limit ($\beta = 0.0$)\cite{IWASAKI}.  Given the difference in
the coupling and the type of fermions used, the overlap between
\cite{IWASAKI} and the present work is unclear.

\section{Acknowledgements}
\label{sec:acknowledgements}
     Helpful conversations with S. Chandrasekharan, A. Gocksch,
S. Gottlieb, F. Karsch, G. Kilcup, W. Lee, A. Mueller, B. Svetitsky,
D. Toussaint and D. Zhu are gratefully
acknowledged.  We are especially indebted to S. Kim, D. Sinclair
and their colleagues at Argonne whose discussions originally
attracted our interest to this subject.

\newpage
\newcounter{tab}
\renewcommand{\thetab}{\Roman{tab}}
\setcounter{tab}{0}
%
%
\refstepcounter{tab}
\begin{table}[h]
\begin{center}
\begin{tabular}{|l|c|c|r|c|c|}
\hline
     &  & & & \multicolumn{2}{c|}{Valence run parameters} \\
\cline{5-6}
\multicolumn{1}{|c|}{$\Delta \tau$} & Start & $\beta$ & Total
$\tau$  &
   $m_{\rm val}$ & Valence $\tau$ \\ \hline
0.0078125 & mix1 & 4.50 & 100  & 0.004  &  50--100 \\
          &      & 4.55 &  25  &  --    &   --     \\
          &      & 4.58 & 100  & 0.004  &  50--100 \\
          &      & 4.60 & 15   &  --    &     --   \\
          &      & 4.65 & 200  & 0.004  & 150--200 \\ \cline{2-6}
         & mix2 & 4.55 & 50   &  --    &    --    \\
          &      & 4.60 & 50   &  --    &    --    \\
          &      & 4.65 & 25   &  --    &    --    \\  \hline
\end{tabular}
\begin{description}
\item{Table \ref{tab:run_4}.}  A list of the parameters for the
runs with $N_t=4$.  The mix1 start was produced by thermalizing
a hot lattice at $\beta=4.5$ for 80 time units and then varying
$\beta$ for 95 time units. The mix2 start was produced by
thermalizing a cold lattice for 70 time units at $\beta=4.5$ and
then varying $\beta$ for 112.5 time units.
\label{tab:run_4}
\end{description}
\end{center}
\end{table}
%
%
\refstepcounter{tab}
\begin{table}[h]
\begin{center}
\begin{tabular}{|l|c|c|r|c|c|}
\hline
     &  & & & \multicolumn{2}{c|}{Valence run parameters} \\
\cline{5-6}
 \multicolumn{1}{|c|}{$\Delta \tau$}& Start & $\beta$ & Total
$\tau$  &
   $m_{\rm val}$ & Valence $\tau$ \\ \hline
0.0078125 & mix & 4.65  & 25 & 0.004  & 0--25   \\
       &     &       &    & 0.010  & 0--25 \\ \cline{3-6}
       &     & 4.68  & 50 & 0.004  & 0--50   \\
       &     &       &    & 0.010  & 0--50 \\ \cline{3-6}
       &     & 4.70  & 50 & 0.004  & 0--50   \\
       &     &       &    & 0.010  & 0--50 \\ \cline{3-6}
       &     & 4.73  & 50 & 0.004  & 0--50    \\
       &     &       &    & 0.010  & 0--50 \\ \cline{2-6}
       &cold & 4.70  & 100& 0.004  & 0--100 \\
       &     &       &    & 0.010  & 0--100 \\ \cline{2-6}
       & hot & 4.70  & 100& 0.004  & 0--100 \\
       &     &       &    & 0.010  & 0--100 \\ \hline
\end{tabular}
\begin{description}
\item[Table \ref{tab:run_6}.] A list of the parameters
for the runs with $N_t=6$.  The mixed start was produced by
thermalizing a hot lattice for 50 time units at $\beta=4.70$
and then varying $\beta$ for 135 time units.
\label{tab:run_6}
\end{description}
\end{center}
\end{table}
%
%
\refstepcounter{tab}
\begin{table}[h]
\begin{center}
\begin{tabular}{|l|c|c|r|c|c|}
\hline
     &  & & & \multicolumn{2}{c|}{Valence run parameters} \\
\cline{5-6}
 \multicolumn{1}{|c|}{$\Delta \tau$}& Start & $\beta$ & Total
$\tau$  &
   $m_{\rm val}$ & Valence $\tau$ \\ \hline
0.0125 & mix & 5.20  & 25  & 0.004  & 0--25     \\
       &     & 5.25  & 50  & 0.004  & 0--50     \\
       &     & 5.28  & 50  & 0.004  & 0--50     \\
       &     & 5.30  & 50  & 0.004  & 0--50     \\
       &     & 5.35  & 50  & 0.004  & 0--50     \\  \hline
0.0078125 & mix & 4.70  & 100 & 0.004  & 25--100   \\
       &     & 4.73  & 125 & 0.004  & 75--125   \\
       &     & 4.75  & 165 & 0.004  & 40--165   \\
       &     & 4.80  & 50  & 0.004  & 25--50    \\  \hline
0.005  & mix & 4.60  & 93.75& 0.004 & 0--93.75  \\
       &     & 4.63  & 62.5& 0.004  & 0--62.5   \\
       &     & 4.65  & 62.5& 0.004  & 0--62.5   \\
       &     & 4.73  & 93.75& 0.004 & 0--93.75  \\  \hline
0.002  & cold& 4.59  & 50  & 0.004  & 0--50     \\ \cline{2-6}
   & mix & 4.55  & 105 & 0.004  & 0--105    \\
       &     & 4.58  & 25  & 0.004  & 0--25     \\
       &     & 4.60  & 25  & 0.004  & 0--25     \\
       &     & 4.62  & 145 & 0.004  & 0--145    \\ \hline
exact  & mix & 4.58  & 75  & 0.004  & 0--75     \\
       &     & 4.60  & 75  & 0.004  & 0--75     \\ \hline
\end{tabular}
\begin{description}
\item[Table \ref{tab:run_8}.] A list of the parameters
for the runs with $N_t=8$.  The mixed start was produced by
thermalizing a cold lattice for 40 time units at $\beta=4.60$ with
$\Delta \tau = 0.0078125$ and then varying $\beta$ for 60 time
units. The $\Delta \tau = 0.005$ runs had a trajectory length of
0.625 time units.
\label{tab:run_8}
\end{description}
\end{center}
\end{table}
%
%
\refstepcounter{tab}
\begin{table}[h]
\begin{center}
\begin{tabular}{|c|l|c|}
\hline
 $N_t$   &\multicolumn{1}{c|}{$\Delta \tau$} & $\beta_c$ \\ \hline
    4        &     0.0078125   &   4.58(1)     \\ \hline
    6        &     0.0078125   &   4.71(1)     \\ \hline
    8        &     0.0125      &   5.29(1)     \\
             &     0.0078125   &   4.73(1)     \\
             &     0.005       &   4.64(1)     \\
             &     0.002       &   4.59(1)     \\
             &     exact       &   4.59(1)     \\ \hline
   16        &     0.0078125   &   4.73(1)     \\
             &     0.005       &   4.62(1)     \\ \hline
\end{tabular}
\begin{description}
\item[Table \ref{tab:betac}.]  Values for $\beta_c$ for
$N_t = 4$, 6, 8 and 16.
\label{tab:betac}
\end{description}
\end{center}
\end{table}
%
%
\refstepcounter{tab}
\begin{table}[h]
\begin{center}
\begin{tabular}{|l|c|c|r|c|l|c|l|}
\hline
       &    &    &   & &  &\multicolumn{2}{c|}  {Valence results}
 \\ \cline{7-8}
\multicolumn{1}{|c|}{$\Delta \tau$}& Start & $\beta $ &
   \multicolumn{1}{c|}
  {$\tau_{\rm eq}$} & Action & \multicolumn{1}{c|}{$\pbp$} &
$m_{\rm val}$ & \multicolumn{1}{c|}{$\pbp_{\rm val}$} \\ \hline
0.0078125&mix1 & 4.50 & 50 & 0.6344(6) & 0.418(2) & 0.004 &
0.410(4) \\
       &       & 4.65 & 50 & 0.4939(1) & 0.0343(3)& 0.004 &
0.0094(4)\\
       \hline
\end{tabular}
\begin{description}
\item[Table \ref{tab:result_4}.] Results for $\pbp$ and the gauge
action for $N_t=4$.
\label{tab:result_4}
\end{description}
\end{center}
\end{table}
%
%
\refstepcounter{tab}
\begin{table}[h]
\begin{center}
\begin{tabular}{|l|c|c|r|c|l|c|l|}
\hline
       &    &    &   & &  &\multicolumn{2}{c|}  {Valence results}
 \\ \cline{7-8}
\multicolumn{1}{|c|}{$\Delta \tau$}& Start & $\beta $ &
   \multicolumn{1}{c|}
  {$\tau_{\rm eq}$} & Action & \multicolumn{1}{c|}{$\pbp$} &
$m_{\rm val}$ & \multicolumn{1}{c|}{$\pbp_{\rm val}$} \\ \hline
0.0078125 & hot  & 4.70 & 50  & 0.5995(4) & 0.378(2)  & 0.004 &
    0.371(2) \\
          &      &      &     &           &           & 0.010 &
    0.375(2) \\ \cline{2-8}
          & cold & 4.70 & 50  & 0.4867(2) & 0.0445(4) & 0.004 &
    0.0126(4) \\
          &      &      &     &           &           & 0.010 &
    0.0302(4) \\ \hline
\end{tabular}
\begin{description}
\item[Table \ref{tab:result_6}.] Results for $\pbp$ and the gauge
action for $N_t=6$.
\label{tab:result_6}
\end{description}
\end{center}
\end{table}
%
%
\refstepcounter{tab}
\begin{table}[h]
\begin{center}
\begin{tabular}{|l|c|c|r|c|l|c|l|}
\hline
       &    &    &   & &  &\multicolumn{2}{c|}  {Valence results}
 \\ \cline{7-8}
\multicolumn{1}{|c|}{$\Delta \tau$}& Start & $\beta $ &
   \multicolumn{1}{c|}
  {$\tau_{\rm eq}$} & Action & \multicolumn{1}{c|}{$\pbp$} &
$m_{\rm val}$ & \multicolumn{1}{c|}{$\pbp_{\rm val}$} \\ \hline
0.0125 & mix & 5.25 & 30  & 0.6908(3) & 0.472(3)  & 0.004 &
0.467(5) \\
       &     & 5.35 & 30  & 0.4168(4) & 0.0272(5) & 0.004
&0.0077(5) \\
       \hline
0.0078125&mix& 4.70 & 70  & 0.6006(3) & 0.378(1)  & 0.004 &
0.372(2) \\
       &     & 4.75 & 100 & 0.4804(4) & 0.0465(3) & 0.004
&0.0131(3)\\
       \hline
0.005  & mix & 4.60 & 62.5& 0.5793(5) & 0.348(2)  & 0.004 &
0.341(3)\\
       &     & 4.73 & 62.5& 0.4838(1) & 0.0493(6) & 0.004
&0.0132(6)\\
       \hline
\end{tabular}
\begin{description}
\item[Table \ref{tab:result_8}.] Results for $\pbp$ and the gauge
action for $N_t=8$.
\label{tab:result_8}
\end{description}
\end{center}
\end{table}
%
\clearpage
\refstepcounter{tab}
\begin{table}[h]
\begin{center}
\begin{tabular}{|l|c|c|r|c|c|}
\hline
     &  & & & \multicolumn{2}{c|}{Valence run parameters} \\
\cline{5-6}
\multicolumn{1}{|c|}{$\Delta \tau$}& Start & $\beta$ &
   Total $\tau$  & $m_{\rm val}$ & Valence $\tau$ \\ \hline
0.0078125 & cold& 4.65 & 342.5 & 0.004 &  0--215  \\
       &     &      &       & 0.010 &  215--342.5 \\ \cline{2-6}
      & hot & 4.65 & 380   & 0.004 &   0--210  \\
       &     &      &       & 0.010 &  210--380 \\ \cline{3-6}
   &     & 5.00 & 40    & 0.010 &   0--40   \\ \cline{2-6}
& mix & 4.71 & 50    &   --     &  --  \\
       &     & 4.73 & 100   & 0.004 &  50--100 \\
       &     & 4.75 & 100   & 0.004 &  50--100 \\ \hline
0.005  & cold& 4.62 & 350   & 0.004 &  0--350  \\ \cline{2-6}
  & hot & 4.62 & 100   & 0.004 &  0--100 \\ \cline{2-6}        &
mix & 4.60 & 50    & 0.004 &  0--50  \\
       &     & 4.62 & 25    & 0.004 &  0--25  \\
       &     & 4.63 & 25    & 0.004 &  0--25  \\
       &     & 4.64 & 50    & 0.004 &  0--50  \\ \hline
\end{tabular}
\begin{description}
\item[Table \ref{tab:run_16}.] A list of the parameters for the
runs with $N_t=16$. The mix start was produced by thermalizing a
cold lattice for 342.5 time units at $\beta=4.65$ with $\Delta \tau
= 0.0078125$ and then varying $\beta$ for 55 time units.
\label{tab:run_16}
\end{description}
\end{center}
\end{table}
%
%
\refstepcounter{tab}
\begin{table}[h]
\begin{center}
\begin{tabular}{|l|c|c|r|c|c|}
\hline
     &  & & & \multicolumn{2}{c|}{Valence run parameters} \\
\cline{5-6}
 \multicolumn{1}{|c|}{$\Delta \tau$}& Start & $\beta$ &
 Total $\tau$  & $m_{\rm val}$ & Valence $\tau$ \\ \hline
0.0078125&cold& 4.60 & 485   & 0.004   &  0--355      \\
       &      &      &       & 0.010   &  355--485    \\
\cline{3-6}
       &      & 4.65 & 865   & 0.004   &  232.5--382.5  \\
       &      &      &       &         &  532.5--765  \\
\cline{5-6}
       &      &      &       & 0.010   &  0--232.5     \\
       &      &      &       &         &  382.5--765   \\
\cline{5-6}
       &      &      &       & 0.025   &  532.5--865   \\
       &      &      &       & 0.050   &  765--865     \\
       &      &      &       & 0.100   &  765--817.5   \\
       &      &      &       & 0.200   &  817.5--865   \\
\cline{3-6}
       &      & 4.70 & 100   & 0.004   &  0--100       \\
\cline{3-6}
       &      & 4.80 & 75    & 0.004   &  0--75        \\
\cline{3-6}
       &      & 4.90 & 100   & 0.004   &  0--100       \\
\cline{3-6}
       &      & 5.00 & 1325  & 0.004   & 1182.5--1325  \\
       &      &      &       & 0.010   & 397.5--1182.5 \\ \hline
\end{tabular}
\begin{description}
\item[Table \ref{tab:run_32}.] A list of the parameters for the
runs with  $N_t=32$. The $\beta=4.60$ run tunneled around $\tau =
250$. \label{tab:run_32}
\end{description}
\end{center}
\end{table}
%
%
\refstepcounter{tab}
\begin{table}[h]
\begin{center}
\begin{tabular}{|l|c|c|r|c|l|c|l|}
\hline
       &    &    &   & &  &\multicolumn{2}{c|}  {Valence results}
 \\ \cline{7-8}
\multicolumn{1}{|c|}{$\Delta \tau$}& Start & $\beta $ &
  \multicolumn{1}{c|}
  {$\tau_{\rm eq}$} & Action & \multicolumn{1}{c|}{$\pbp$} &
$m_{\rm val}$ & \multicolumn{1}{c|}{$\pbp_{\rm val}$} \\ \hline
0.0078125&cold&4.65 & 100 & 0.4961(1) & 0.0711(4) & 0.004 &
0.0259(5) \\
       &    &     &     &           &           & 0.010 & 0.0541(5)
\\
  \cline{2-8}
         &hot&4.65& 100 & 0.6111(2) & 0.3953(4) & 0.004 &
0.3900(11)\\
       &    &     &     &           &           & 0.010 & 0.3933(6)
\\
  \hline
0.005  &cold&4.62 & 100 & 0.5006(1) & 0.0829(6) & 0.004 & 0.0399(8)
\\
  \cline{2-8}
       &hot &4.62 & 75  & 0.5631(3) & 0.314(1)  & 0.004 & 0.305(2)
\\
  \hline
\end{tabular}
\begin{description}
\item[Table \ref{tab:result_16}.]
Results for $\pbp$ and the action for $N_t=16$.
\label{tab:result_16}
\end{description}
\end{center}
\end{table}
%
%
\refstepcounter{tab}
\begin{table}[h]
\begin{center}
\begin{tabular}{|l|c|c|r|c|l|c|l|}
\hline
       &    &    &   & &  &\multicolumn{2}{c|}  {Valence results}
 \\ \cline{7-8}
 \multicolumn{1}{|c|}{$\Delta \tau$}& Start & $\beta $ &
   \multicolumn{1}{c|}
  {$\tau_{\rm eq}$} & Action & \multicolumn{1}{c|}{$\pbp$} &
$m_{\rm val}$ & \multicolumn{1}{c|}{$\pbp_{\rm val}$} \\ \hline
0.0078125&cold&4.60&325 & 0.62031(6) & 0.4075(3) & 0.004 &
0.4038(16) \\
 & &     &      &           &          & 0.010 &0.4053(4) \\
\cline{3-8}
 & &4.65 &382.5 &0.49552(6) &0.0687(1) & 0.004 &0.0234(3) \\  & &
   &      &           &          & 0.010 &0.0506(3) \\  & &     &
    &           &          & 0.025 &0.0983(1) \\  & &     &      &
         &          & 0.050 &0.1532(2) \\  & &     &      &
   &          & 0.100 &0.2255(2) \\  & &     &      &           &
        & 0.200 &0.3390(2) \\ \cline{3-8}
 & &5.00 &250   &0.45125(2) &0.03539(3)& 0.004 &0.00965(9)\\  & &
   &      &           &          & 0.010 &0.02386(4)\\ \hline
\end{tabular}
\begin{description}
\item[Table \ref{tab:result_32}.]
Results for $\pbp$ and the action for $N_t=32$. The
$\beta=4.60$ run tunneled at about $\tau=250$.
\label{tab:result_32}
\end{description}
\end{center}
\end{table}
%
%
\refstepcounter{tab}
\begin{table}[h]
\begin{center}
\begin{tabular}{|c|c|c|c|c|}
 \hline($J^{PC}$)  &$\beta=4.65$ strong.
                             &$\beta=4.65$ weak
                                        &$\beta=5.00$
                                                   &$\beta=5.7$
                                                              \\ \hline
$\pi (0^{-+})$     &0.297(1) &0.378(2)  &0.405(3)  &0.293(2)  \\ \hline
$\pi_2 (0^{-+})$   &1.60(8)  &0.471(7)  &0.434(10) &0.333(3)  \\ \hline
$\sigma (0^{++})$  &-        &0.465(3)  &0.415(4)  &0.487(12) \\ \hline
$\rho (1^{--})$    &1.41(1)  &0.522(7)  &0.484(7)  &0.455(8)  \\ \hline
$\rho_2 (1^{--})$  &1.64(2)  &0.521(4)  &0.490(5)  &0.452(7)  \\ \hline
$A_1 (1^{+-})$     &-        &0.582(11) &0.491(7)  &0.594(22) \\ \hline
$N({1 \over 2}^+)$ &2.29(7)  &0.872(10) &0.807(7)  &0.685(10) \\ \hline
$N'({1\over 2}^-)$ &-        &0.896(13) &0.810(6)  &0.833(38) \\ \hline
$B_1 (1^{++})$     &-        &0.586(26) &0.512(8)  &0.596(28) \\ \hline
\end{tabular}

\begin{description}
\item[Table \ref{tab:masses_sea}.] Hadron masses for quark mass
$m_{val}a=m_{sea}a=0.015$.  The second column was obtained on a
$16^4$ lattice beginning with a hot start and the third and fourth
on $16^3 \times 32$ lattices with a cold start.  For reference, the
right column lists the results of our earlier $N_f=2$, $16^3 \times 32$
calculation with
$m_{sea}a=m_{val}a=0.015$\cite{HONG,HADRON_SHORT,HADRON_LONG}.
\label{tab:masses_sea}
\end{description}
\end{center}
\end{table}
%
%
\refstepcounter{tab}
\begin{table}[h]
\begin{center}
\begin{tabular}{|c|c|c|c|c|c|}
\hline ($J^{PC}$) & $\beta=4.65$ strong

                      &\multicolumn{4}{c|}{$\beta=4.65$ weak}
                           \\
\hline
          & $m_{val}a=0.01$&$=0.004$&$=0.01$ &$=0.025$&$=0.05$  \\
\cline{2-6}
$\pi (0^{-+})$
    &0.243(1) &0.257(4)  &0.328(2) &0.465(4) &0.619(2)
\\ \hline
$\pi_2 (0^{-+})$
    &1.65(12)  &0.369(32) &0.420(6) &0.564(5) &0.767(7)        \\
\hline
$\sigma (0^{++})$
    &-           &0.284(3)  &0.391(3) &0.587(5) &0.790(6)        \\
\hline
$\rho (1^{--})$
    &1.40(2)  &0.404(11) &0.465(9) &0.632(6) &0.813(4)
\\ \hline
$\rho_2 (1^{--})$
    &1.63(2)  &0.407(11) &0.468(7) &0.634(7) &0.817(7)          \\
\hline
$A_1 (1^{+-})$
    &-           &0.428(28) &0.511(14) &0.751(40) &0.951(26)
 \\ \hline
$N ({1 \over 2}^+)$
    &2.40(15)  &0.747(30) &0.818(21) &0.992(15) &1.246(8)        \\
\hline
$N' ({1 \over 2}^-)$
    &-          &0.744(34) &0.836(28) &1.043(31) &1.304(29)
\\ \hline
$B_1 (1^{++})$
    &-         &0.429(21) &0.511(24) &0.765(45) &1.045(73)       \\
\hline
\end{tabular}
\begin{description}
\item[Table \ref{tab:masses_val_4_65}.] Hadron
masses calculated with a variety of valence quark masses.  The
masses quoted in the second column were obtained on a $16^4$
lattice beginning with a hot start while those in columns
three through six came from a cold start using a $16^3 \times
32$ lattice.
\label{tab:masses_val_4_65}
\end{description}
\end{center}
\end{table}
%
%
\refstepcounter{tab}
\begin{table}[h]
\begin{center}
\begin{tabular}{|c|c|c|}
\hline ($J^{PC}$) &\multicolumn{2}{c|}{$\beta = 5.00$} \\
\hline
          & $m_{val}a=0.004$ & $m_{val}a=0.01$ \\
\cline{2-3}
$\pi (0^{-+})$
    &0.386(9) &0.389(3)           \\ \hline
$\pi_2 (0^{-+})$
    &0.500(108)       &0.422(27)       \\ \hline
$\sigma (0^{++})$
    &0.384(9)         &0.394(5)       \\ \hline
$\rho (1^{--})$
    &0.474(6) &0.465(10)           \\ \hline
$\rho_2 (1^{--})$
    &0.485(8) &0.481(8)         \\ \hline
$A_1 (1^{+-})$
    &0.475(6)  &0.469(10)        \\ \hline
$N ({1 \over 2}^+)$
    &0.783(22) &0.785(8)       \\ \hline
$N' ({1 \over 2}^-)$
   &0.789(28) &0.786(7)          \\ \hline
$B_1 (1^{++})$
  &0.487(8) &0.491(11)        \\ \hline
\end{tabular}
\begin{description}
\item[Table \ref{tab:masses_val_5_00}.] Hadron
masses for two different valence quark masses on a $16^3 \times 32$
lattice.
\label{tab:masses_val_5_00}
\end{description}
\end{center}
\end{table}
\clearpage
\newpage
\noindent
{\bf \LARGE Figure Captions}
\begin{figure}[h]
\caption[fig:phase]  {A phase diagram in the $\beta-N_t$
plane for eight-flavor QCD in infinite spatial volume consistent with
the results presented here.  $N_t$ is the temporal extent of the
lattice and $\beta=6/g^2$ is the lattice-coupling strength.
The solid line, becoming vertical for $N_t \ge 8$
locates a ``zero-temperature", first-order transition---a lattice
artifact.  The dashed line suggests a possible, continuum
finite-temperature transition that occurs in the weak-coupling phase.
The system shows chiral symmetry to the right of and below this dashed
line while we speculate that chiral symmetry will be spontaneously
broken to the left of and above this line.  The solid squares label
parameter values where we have performed simulations, while the open
squares locate critical values.}  \label{fig:phase}
\end{figure}
\begin{figure}[h]
\caption[fig:2f_extrap]  {The quark mass dependence seen
for $\pbp$ and $m_\pi^2$ in earlier two-flavor calculations.  The
$ma=0.01$, 0.015, 0.02 and 0.025 points are calculations properly
including the effects of dynamical quarks\cite{HADRON_SHORT}, while
the $ma=0.004$ point is obtained using that value in the explicit
quark propagators but the value $ma=0.01$ in the fermion
determinant.  The lines correspond to the fits in
Eq.~(\ref{eq:2f_fit}).}
\label{fig:2f_extrap}
\end{figure}
\begin{figure}[h]
\caption[fig:mix1_nt4]  {Generation of a mixed-phase
configuration from a hot start for $N_t=4$.  The lower,
cold-start trajectory establishes the value of $\pbp$ for the
chirally symmetric phase while the upper trajectory both gives a
value of $\pbp$ in the symmetry broken phase and with a subsequent
tuning of $\beta$ becomes our candidate ``mixed'' phase.}
\label{fig:mix1_nt4}
\end{figure}
\begin{figure}[h]
\caption[fig:mix1_nt4_beta]  {Determining $\beta_c$ for a
mixed-phase configuration generated from a hot starting
lattice with $N_t=4$.  From this figure we conclude that the
critical value of $\beta$ is $\beta_c=4.58(1)$.}
\label{fig:mix1_nt4_beta}
\end{figure}
\begin{figure}[h]
\caption[fig:mix2_nt4_beta]  {Determining $\beta_c$ for a
mixed-phase configuration generated from a cold starting
lattice with $N_t=4$.  From this figure we conclude that the
critical value of $\beta$ is $\beta_c=4.58(1)$.}
\label{fig:mix2_nt4_beta}
\end{figure}
\begin{figure}[h]
\caption[fig:mix_nt8_beta]  {Determining $\beta_c$ for a
mixed-phase configuration generated from a cold starting
lattice with $N_t=8$.  From this figure we conclude that the
critical value of $\beta$ is $\beta_c=4.73(1)$.}
\label{fig:mix_nt8_beta}
\end{figure}
\begin{figure}[h]
\caption[fig:betac]  {$\beta_c$ versus $\Delta \tau$  for
$N_t=8$.  The curve is a quadratic fit to the points
$\Delta \tau = 0.002$, 0.005 and 0.0078125.} \label{fig:betac}
\end{figure}
\begin{figure}[h]
\caption[fig:468extrap]  {Extrapolation of $\pbp$  for
$N_t=4$, 6 and 8 as a function of valence quark mass, $m_{val}$.
The fits are forced through the origin.}
\label{fig:468extrap}
\end{figure}
\begin{figure}[h]
\caption[fig:4_65_two_phase]  {The evolution of $\pbp$ for two
independent Monte Carlo runs on a $16^4$ lattice at $\beta=4.65$.
The upper curve represents a run begun with a hot start while the
lower curve began with a cold start.} \label{fig:4_65_two_phase}
\end{figure}
\begin{figure}[h]
\caption[fig:4_60_tunnel]  {The evolution of $\pbp$
{}from a cold start on a $16^3 \times 32$ lattice with
$\beta=4.60$.  We interpret the jump seen at $\tau \approx 250$ as
tunneling from the metastable, weak-coupling phase to the stable,
strong-coupling phase.} \label{fig:4_60_tunnel}
\end{figure}
\begin{figure}[h]
\caption[fig:mix_nt16_beta]  {Determining $\beta_c$ for a
mixed-phase configuration generated from a cold starting
lattice with $N_t=16$.  From this figure we conclude that the
critical value of $\beta$ is $\beta_c=4.73(1)$.}
\label{fig:mix_nt16_beta}
\end{figure}
\begin{figure}[h]
\caption[fig:4_65_strong_extrap]  {Linear fits to $\pbp$
and $m_\pi^2$ in the strong-coupling phase at $\beta=4.65$ on a
$16^4$ lattice.  The $m_{\pi}^2$ fit is forced through the origin
and has $\chi^2/dof = 0.1/1$.} \label{fig:4_65_strong_extrap}
\end{figure}
\begin{figure}[h]
\caption[fig:4_65_weak_extrap]  {$\pbp$
and $m_\pi^2$ in the weak-coupling phase at $\beta=4.65$ on a
$16^3 \times 32$ lattice. The line is a linear fit to $m_\pi^2$.}
\label{fig:4_65_weak_extrap}
\end{figure}
\begin{figure}[h]
\caption[fig:4_65_weak_mass]  {Values $m_\pi$,
$m_\sigma$, $m_\rho$, $m_{A_1}$, $m_N$ and $m_{N'}$ plotted versus
$m_{val}a$ for a $16^3 \times 32$ lattice in the weak-coupling
phase with $\beta=4.65$. The lines shown correspond to the fits in
Eq.~(\ref{eq:4_65_linear_fit}).} \label{fig:4_65_weak_mass}
\end{figure}
\begin{figure}[h]
\caption[fig:5_00_extrap] {$\pbp$ and $m_\pi^2$ plotted as
a function of $m_{val}a$ for $\beta=5.00$ on a $16^3 \times 32$ lattice.
The straight lines
shown are least squares fits to the three mass values
$m_{val}a= 0.004$, 0.01 and 0.015.  Both quantities show the
behavior expected in a chirally symmetric phase.}
\label{fig:5_00_extrap}
\end{figure}
\begin{figure}[h]
\caption[fig:squash] {A sketch of the $\beta_c$
dependence of $N_t$ for the finite-temperature QCD phase transition
for $N_f \le 4$.  The area between the vertical, dotted lines
represents the region of $\beta$ for which infinite space-time
volume systems show cross-over behavior from strong to weak coupling.
The increased slope of the $N_t$-versus-$\beta_c$ curve in this
region displays the well-established non-scaling behavior seen for
$T_c\,a$ in this region.
The dashed line has a slope predicted by the perturbative
renormalization group.} \label{fig:squash}
\end{figure}
\begin{figure}[h]
\caption[fig:n_f-summary]  {Values of $N_t$ versus
$\beta_c$ are plotted for zero\cite{000}, two\cite{222},
four\cite{444} and eight flavors.  The dashed lines have slopes
predicted by the perturbative renormalization group and have been
located to show the possible weak-coupling behavior of the adjacent
curve.} \label{fig:n_f-summary}
\end{figure}

\begin{thebibliography}{99}
\bibitem{KOGUT_PRL} J. B. Kogut, J. Polonyi, H. W. Wyld and D. K.
Sinclair, Phys.~Rev.~Lett. {\bf 54}, 1475 (1985).
%
\bibitem{KOGUT_NP} J. B. Kogut and D. K. Sinclair, Nucl.~Phys.~{\bf
B295}[FS21], 465 (1988).
%
\bibitem{FUKUGITA} M. Fukugita, S. Ohta and A. Ukawa,
Phys.~Rev.~Lett.~{\bf 60}, 178 (1988).
%
\bibitem{OHTA} S. Ohta and S. Kim, Phys.~Rev.~{\bf 44}, 504 (1991)
and Columbia University preprint, in preparation.
%
\bibitem{GOTTLIEB}  For a recent of review of finite-temperature
QCD with $N_f=2$, 3 and 4 see S. Gottlieb, Nucl.~Phys.~B (Proc.~Suppl.)
{\bf 20}, 247 (1991).
%
\bibitem{DETAR_KOGUT} C. DeTar and J. B. Kogut,
Phys.~Rev.~Lett.~{\bf 59} 399 (1987); Phys.~Rev.~D {\bf 36},  2828
(1987).
%
\bibitem{R_ALGORITHM} S. Gottlieb {\it et al.}, Phys.~Rev.~D {\bf
35}, 2531 (1987).
%
\bibitem{256_NODE} A. Vaccarino, Nucl.~Phys.~B (Proc. Suppl.) {\bf
17}, 421 (1990); N. H. Christ, Nucl.~Phys.~B (Proc. Supp.) {\bf
17}, 267 (1990).
%
\bibitem{DUANE} S. Duane {\it et al.}, Phys.~Lett.~B {\bf 195}, 216
(1987).
%
\bibitem{OPERATORS}
K.~C.~Bowler, Nucl.~Phys.~B {\bf 284,} 299 (1987).
%
\bibitem{HONG} H. Chen, Nucl.~Phys.~B (Proc.~Suppl.) {\bf 20}, 370
(1991).
%
\bibitem{HADRON_SHORT} F. R. Brown {\it et al.},
Phys.~Rev.~Lett.~{\bf 67}, 1062 (1991).
%
\bibitem{HADRON_LONG} H. Chen, {\it Hadron Spectrum from Lattice
QCD with a 16 Gigaflop Parallel Processor}, Columbia University
Ph.D. thesis (1991); F. R. Brown, {\it et al.}, in preparation.
%
\bibitem{MIXED} For an early application of this technique see
M.~Creutz, L.~Jacobs and C.~Rebbi, Phys.~Rev.~Lett.~{\bf 42}, 1390
(1979).
%
\bibitem{PURE_GAUGE_TRANSITION} S. A. Gottlieb {\it et al.},
Phys.~Rev.~Lett.~{\bf 55}, 1958 (1985); N. H. Christ and A. E. Terrano,
Phys.~Rev.~Lett.~{\bf 56}, 111 (1986);  N. H. Christ, in {\it
Proceedings of the Rice Meeting} Billy Bonner and Hannu Miettinen
eds., pg.~780, World Scientific (1990).
%
\bibitem{THERMO_NF4} F.~R.~Brown, H.~Chen, N.~H.~Christ, Z.~Dong,
W.~Shaffer, L.~I.~Unger and A.~Vaccarino, Phys.~Lett.~B {\bf 251},
181 (1990).
%
\bibitem{SU(2)_MIX_ACTION} G.~Bhanot and M.~Creutz, Phys.~Rev.~D
{\bf 24}, 3212 (1981).
%
\bibitem{SU(3)_SCALE_VIOLATION} A. D. Kennedy {\it et al.},
Phys.~Rev.~Lett.~{\bf 54}, 87 (1985).
%
\bibitem{000} For $N_t=2$: A.~D.~Kennedy, J.~Kuti, S.~Meyer and
B.~J.~Pendelton, Phys.~Rev.~Lett.~{\bf 54}, 87 (1985). For $N_t=4$:
F.~R.~Brown, et.~al., Phys.~Rev.~Lett.~{\bf 61}, 2058 (1989).
For $N_t = 6$, 8 and 10: N.~H.~Christ and H.~Q.~Ding,
Phys.~Rev.~Lett.~{\bf 60}, 1367 (1988).  For $N_t=12$ and 14:
N.~H.~Christ and A.~E.~Terrano, Phys.~Rev.~Lett.~{\bf 56}, 111 (1986).
For $N_t=16$: N.~H.~Christ, Nucl.~Phys.~B (Proc.~Suppl .) {\bf 17},
267 (1990).
%
\bibitem{222} For $N_t=4$: F.~R.~Brown, et.~al.,
Phys.~Rev.~Lett.~{\bf 65}, 2491 (1990); A.~Vaccarino, Ph.D.~Thesis,
Columbia University, unpublished (1991). For $N_t=6$: C.~Bernard,
et.~al., Indiana University preprint IUHET-210.  For $N_t=8$:
S.~Gottlieb, et.~al., Proceedings of Lattice '91, Tsukuba, Japan,
to appear in Nucl.~Phys.~B (Proc.~Suppl.).
%
\bibitem{444} For $N_t=4$ and 6: A.~Vaccarino, Nucl.~Phys.~B
(Proc.~Suppl.) {\bf 17}, 441 (1990); F.~R.~Brown, et.~al.,
Phys.~Lett.~B {\bf 251}, 181 (1990); F.~Butler, Ph.D.~Thesis,
Columbia University, unpublished (1990); A.~Vaccarino, Ph.D.~Thesis,
Columbia University, unpublished (1991).
%
\bibitem{T'HOOFT} G.~'t~Hooft, {\em Recent Developments in Gauge
Theories}, G.~'t~Hooft ed., Plenum Press, 1980.
%
\bibitem{BANKS&ZAKS} T. Banks and A. Zaks, Nucl.~Phys.~B {\bf 196},
189 (1982).
%
\bibitem{IWASAKI} Y.~Iwasaki, talk presented at Lattice '91, Tsukuba,
Japan, November, 1991.
%
\end{thebibliography}
\end{document}